\documentclass[a4paper,twocolumn,prl,showpacs,amsmath,amssymb]{revtex4}

\usepackage[T1]{fontenc}
\usepackage[latin1]{inputenc}
\usepackage{color}
\usepackage{vmargin}
\usepackage[dvips]{graphicx}
\usepackage{txfonts}

\setmarginsrb{20mm}{20mm}{20mm}{15mm}{15pt}{11mm}{0pt}{11mm}%

\renewcommand{\vec}[1]{\boldsymbol{#1}}

\newcommand{\curl}{\vec{\nabla}\times}

\newcommand{\f}[2]{\frac{#1}{#2}}

\newcommand{\rey}{\mathrm{Re}}
\newcommand{\reym}{\mathrm{Rm}}
\newcommand{\rom}{R_\Omega}

\newcommand{\difft}[1]{\partial_t #1}

\begin{document}

\title{A self-sustaining nonlinear dynamo process in Keplerian shear flows}

\author{F. Rincon}
\email{F.Rincon@damtp.cam.ac.uk}

\author{G. I. Ogilvie}
\author{M. R. E. Proctor}

\affiliation{Department of Applied Mathematics and Theoretical Physics,
University of Cambridge,\\ Centre for Mathematical Sciences, Wilberforce
Road, Cambridge CB3 0WA, United Kingdom}

\date{\today}
%\nocite{*}

\begin{abstract}
A three-dimensional nonlinear dynamo process is identified in rotating
plane Couette flow in the Keplerian regime. It is analogous to the
hydrodynamic self-sustaining process in non-rotating shear flows and
relies on the magneto-rotational instability of a toroidal
magnetic field. Steady nonlinear solutions are computed numerically for
a wide range of magnetic Reynolds numbers but are restricted to low
Reynolds numbers. This process may be important to explain the
sustenance of coherent fields and turbulent motions in Keplerian
accretion disks, where all its basic ingredients are present.
\end{abstract}


\maketitle

The most natural explanation for the efficient outward angular momentum
transport inferred from observed accretion luminosities in accretion disks
is that these objects are turbulent \citep{balbus98}. In
magnetized disks, turbulence is likely triggered
by the magneto-rotational instability  \citep[MRI,
see][]{velikhov59}. When a mean magnetic field
with non-zero net flux perpendicular to the disk plane is imposed, the
instability takes the form of a two-dimensional channel flow which
breaks down into three-dimensional MHD turbulence \citep{hawley95} as a
result of secondary Kelvin-Helmholtz instabilities
\citep{goodman94}. This simple field configuration, however, may not be
relevant to all accretion disks. Recent observations have for
instance demonstrated that the magnetic fields in the innermost regions of
some disks are probably not created by an external object such
as the central accreting body but are  intrinsic to the disk
\citep{donati05}. Their precise origin, however, remains largely unknown. 
Understanding the nonlinear saturation of the MRI in the absence of
an externally imposed magnetic field or more generally when there is no net
magnetic flux threading through the disk proves to be a complicated task
because one needs to explain in that case how the magnetic field, whose
presence within the disk is permanently required for the generation of
MHD turbulence, can be sustained  against dissipation.
Local numerical disk simulations in the absence of a net magnetic flux 
\citep{branden95,fleming00} suggest that a dynamo process
possibly relying on the MRI of toroidal magnetic fields
\citep{ogilvie96} may be at work, but a detailed theoretical
understanding of such a scenario is still lacking.

In this Letter, we report the discovery of self-sustaining 
dynamo action in magnetized, spanwise rotating plane Couette flow (PCF)
in the Keplerian regime (linearly stable from the purely hydrodynamic
point of view) characteristic of accretion disks. The phenomenology of this
magnetohydrodynamic (MHD) process is analogous to that of the
self-sustaining process (SSP) thought to be responsible for
the transition to turbulence in hydrodynamic 
shear flows \citep{hamilton95,waleffe98}. It
relies on three fundamental physical effects:
\begin{enumerate}
\item linear amplification of zero-net-flux toroidal (azimuthal in 
  the accretion disk terminology) magnetic field induced by the
  distortion of a weak poloidal (radial and
  vertical) seed field by the background shear (differential
  rotation) ; %. %This linear effect is often referred to as the
%  $\Omega$-effect in mean-field dynamo theory ;
\item three-dimensional linear instability (MRI) of the toroidal
  magnetic field ;
\item regeneration of the poloidal field owing to
  the nonlinear feedback of the MRI modes.
\end{enumerate}
Even though linear processes are essential to the mechanism, either
via transient linear growth or linear instability,
the whole process is fundamentally nonlinear since the self-sustaining
loop cannot be closed without nonlinear feedback. 
This notably means that there is no kinematic regime for the dynamo,
which is therefore subcritical: an initial finite-amplitude,
zero-net-flux magnetic field disturbance is required for 
the dynamo to operate. This contrasts with the externally imposed
field problem in which infinitesimal disturbances grow exponentially
due to the MRI. To the best of our knowledge, this is the first
instance of an explicit nonlinear subcritical dynamo solution.
%They also
%illustrate a very generic mechanism for the sustenance of
%large-scale, zero net flux magnetic  fields in Keplerian shear flows. 
We first describe the three steps of the process, 
construct steady nonlinear solutions for Keplerian
PCF at various magnetic and kinetic Reynolds numbers and finally
discuss the relevance of our results to accretion disk theory.

We consider PCF for an incompressible fluid with unit
density, constant kinematic viscosity $\nu$ and magnetic diffusivity $\eta$. 
The flow is driven by two counter-moving rigid, no-slip, perfectly
conducting walls located at $y=\pm d$ and is rotating at a constant 
rate $\Omega$ along the spanwise (vertical) $z$-axis perpendicular to the 
linear background
shear flow $\vec{V}_B\,(y)=Sy\,\vec{e}_x$. It is taken to be spatially
periodic in both  streamwise (toroidal, $x$)
and spanwise $(z)$ directions, with periods $L_x=2\pi d/\alpha$ and
$L_z=2\pi d/\beta$.  Using $1/S$ as a time unit and 
the channel half-width $d$ as a length unit, we define a Reynolds
number $\rey=Sd^2/\nu$ and a magnetic Reynolds number $\reym=Sd^2/\eta$. 
This configuration represents an idealized local model of
a differentially rotating Keplerian accretion disk 
provided that the rotation number $\rom=-2\Omega/S$ equals $-4/3$ 
(anticyclonic Rayleigh-stable rotation). Owing to PCF symmetries,
we look for nonlinear three-dimensional \textit{steady}
solutions of the incompressible MHD equations (Navier-Stokes equation
with a Lorentz force and induction equation) for magnetic 
and velocity field perturbations $\vec{b}$ and $\vec{v}$ of 
Keplerian PCF. We first consider step 1 of the process. 
We define a streamfunction $\psi(y,z,t)$ and a flux function
$\chi(y,z,t)$ to describe the mean poloidal fields 
$\overline{\vec{v}}_{\mathrm{p}}=\curl{\left(\psi\vec{e}_x\right)}$
and
$\overline{\vec{b}}_{\mathrm{p}}=\curl{\left(\chi\vec{e}_x\right)}$,
where overbars stand for $x$-averaging, and denote the $x$-dependent
part of the fields by $\vec{v'}$ and $\vec{b'}$.
%=\vec{v}-\vec{\overline{v}}$ and $\vec{b'}=\vec{b}-\vec{\overline{b}}$).
The toroidal and poloidal components of the $x$-averaged induction equation read
\begin{equation}
\label{eq:induc1}  
\difft{\overline{b}_x}=\overline{b}_y+\vec{e}_x\cdot\overline{\curl{\left(\vec{v}\times\vec{b}\right)}}+\f{1}{\reym}\Delta\,\overline{b}_x~,
\end{equation}
\begin{equation}
 \label{eq:induc2}  
 \difft{\,\chi}+\f{\partial\,\psi}{\partial z}\f{\partial\,\chi}{\partial y}
-\f{\partial\,\psi}{\partial y}\f{\partial\,\chi}{\partial z}
=\overline{\left(\vec{v'}\times\vec{b'}\right)}\cdot\vec{e}_x+\f{1}{\reym}\Delta\,\chi
 %\difft{\,\chi}-\f{\partial\,(\psi,\chi)}{\partial\,
 %(y,z)}=\overline{\left(\vec{v'}\times\vec{b'}\right)}\cdot\vec{e}_x+\f{1}{\reym}\Delta\,\chi
 \end{equation}
% \end{equation}
% \begin{equation}
% \label{eq:induc2}  
% \difft{\overline{b}_y}=\diffz{\,\overline{\left(\vec{v}\times\vec{b}\right)}\cdot\vec{e}_x}+\f{1}{\reym}\Delta\,\overline{b}_y~.
% \end{equation}
Equation~(\ref{eq:induc1}) has a linear induction source term $\overline{b}_y$
due to the presence of the background shear, which 
makes it possible to generate an $O(1)$ toroidal magnetic field 
from a weak $O(1/\reym)$ poloidal field (this results mathematically
from the non-normality of the  linear operator).
This axisymmetric process for two-dimensional, three-component (2D-3C)
fields is often referred to as the $\Omega$-effect in
dynamo theory and is the MHD analog of the algebraic
amplification  of streaks (streamwise velocity field)  in
non-rotating hydrodynamic shear flows, known as the lift-up effect in
the fluid dynamics literature \citep{landahl80}.
Unless some three-dimensional nonlinear mechanism regenerates the
poloidal field, this linear process can however only be
transient: for purely two-dimensional configurations, Eq.~(\ref{eq:induc2})
is a simple advection-diffusion equation with no source term for $\chi$,
which must therefore decay resistively on an $O(\reym)$ timescale.
This is just a restatement of Cowling's theorem for an axisymmetric
($x$-independent here) system. To obtain a SSP, we must therefore
consider the $x$-dependent,
three-dimensional instabilities of the transiently amplified toroidal
component of the 2D-3C field (step 2) and their nonlinear
feedback in the poloidal equation (step 3). Three-dimensional,
nonlinear steady solutions are possible only if the nonlinear
interaction term on the r.h.s. of Eq.~(\ref{eq:induc2}) has the ability
to regenerate the original seed poloidal field. For the hydrodynamic SSP, the
three-dimensional instabilities are inflectional instabilities of the
spanwise-modulated finite-amplitude streaks. In the Keplerian MHD
problem, a natural instability candidate is a
three-dimensional MRI of the $O(1)$ toroidal magnetic field. 

We use a three-dimensional nonlinear continuation code based on
Newton iteration to test this  scenario quantitatively.
The code  is similar to those used  to study hydrodynamic SSPs
\citep{waleffe98,wedin04} and has been tested extensively with
standard nonlinear problems \citep{rincon07}.
Collocation on a Gauss-Lobatto grid is used in 
$y$ and a Fourier representation (with dealiasing) is used in
$x$ and $z$. The Newton solver relies on the linear algebra library
LAPACK to solve real systems with $O(20~000)$ unknowns.
An iterative generalized eigenvalue problem solver from the ARPACK library
is used to address 2D linear stability problems. The convergence of 
our solutions ($\sim 10^{-8}$ for energy) has been thoroughly checked
by looking at Chebyshev and Fourier spectra and by comparing results
obtained at different resolutions.

\begin{figure}[h]
\resizebox{\hsize}{!}{
\includegraphics[height=7.cm]{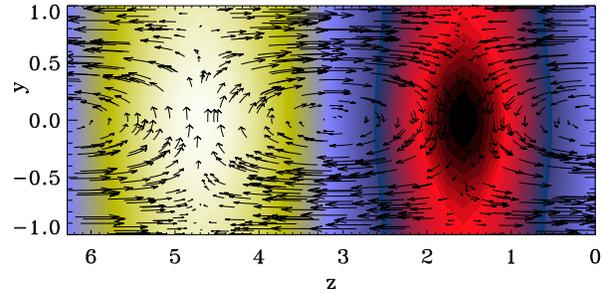}}
  \caption{2D-3C magnetic configuration induced by the
   artificial toroidal EMF~(\ref{eq:emfx}), $\beta=1$ and perfectly
   conducting boundary conditions. $b_x$ is represented on a color scale
   (black to white from -max to max) and
   $(b_y,b_z)$ by arrows. $\reym=750$, $\rey=10$, $A=1.5$,
   $\mathrm{max}(b_x)=1.006$ and $\mathrm{max}(b_y)=0.00195$. $(N_y,N_z)=(32,32)$.}
  \label{fig1}
\end{figure}

We adopt a forcing strategy \citep{waleffe98}, which first consists in 
artificially forcing nonlinear steady 2D-3C solutions and in
computing their three-dimensional instability modes. One must then check 
that the nonlinear feedback of these modes can take over the forcing 
to obtain three-dimensional nonlinear unforced solutions. For the MHD
problem, an appropriate forcing is a toroidal electromotive force (EMF)
\begin{equation}
  \label{eq:emfx}
  \mathrm{EMF}_x\,(y,z)=\f{A}{\beta\,\reym^2}\,\cos\left(\f{\pi\,y}{2}\right)\cos\beta z,
\end{equation}
applied to Eq.~(\ref{eq:induc2}) with $A=O(1)$, 
which compensates for the resistive decay of an $O(1/\reym)$ 
poloidal magnetic field that in turn generates an $O(1)$ toroidal 
field. Fig.~\ref{fig1}
depicts a 2D-3C magnetic field solution of the forced nonlinear MHD
equations obtained with our Newton solver.  

\begin{figure}[h]
\resizebox{\hsize}{!}{
\includegraphics[height=6.cm]{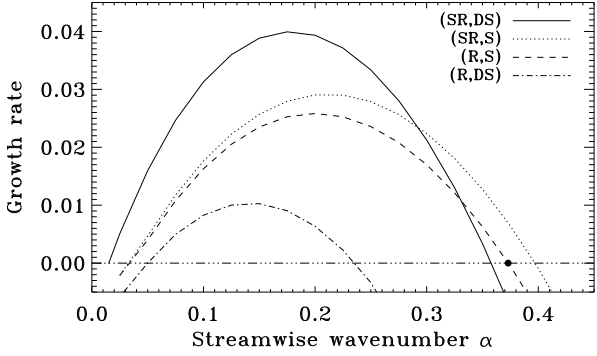}}
  \caption{Growth rates of the most unstable three-dimensional MRI
     eigenmodes for the 2D-3C configuration with $O(1)$ toroidal
     field shown in Fig.~\ref{fig1}, as a function of the streamwise wavenumber
     $\alpha$. The legend indicates the symmetries of  each mode.}
% Full line: (SR,DS) mode. 
%      the dotted line to a (SR,S) mode,  the dashed line to a
%      (R,S) mode and the dash-dotted line to a (R,DS) mode.}}
  \label{fig2}
\end{figure}

\begin{figure}[h]
\resizebox{\hsize}{!}{
\includegraphics[height=6.cm]{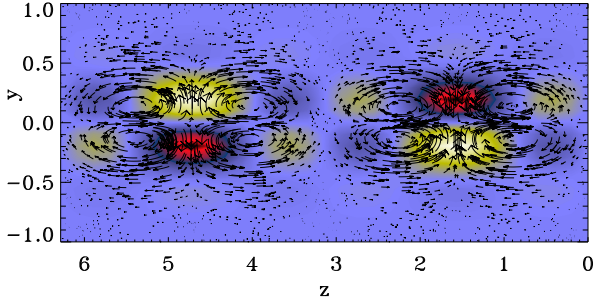}}
  \caption{Magnetic field in the $x=0$ plane of the near marginal (R,S) MRI
    eigenmode with $\alpha=0.375$, depicted by a black dot in
    Fig.~\ref{fig2}. Same representation and parameters as in Fig.~\ref{fig1}.}
  \label{fig3}
\end{figure}

\noindent We then consider the stability of forced 2D-3C MHD flows
   with a dominant $O(1)$ toroidal magnetic field to infinitesimal perturbations
   with  $\exp\left(i\alpha x\right)$ dependence, and compute the
   associated  MRI eigenmodes with our linear eigenvalue 
   solver (the Alfv\'en continuum is removed here owing to viscous and
   resistive effects). For the forcing term~(\ref{eq:emfx}) steady
   solutions have three symmetries which are used to reduce
   computational costs: first, they either have reflect (R)
%    \begin{equation}
% (v_x,v_y,v_z)(x,y,L_z/4+z)\rightarrow (v_x,v_y,-v_z)(x,y,L_z/4-z)
%    \end{equation}
% or shift-and-reflect symmetry
%    \begin{equation}
% (v_x,v_y,v_z)(x,y,L_z/4+z)\rightarrow (v_x,v_y,-v_z)(x+L_x/2,y,L_z/4-z)~,
%    \end{equation}
% a shift-and-rotate symmetry
%    \begin{equation}
% (v_x,v_y,v_z)(x,y,z)\rightarrow (-v_x,-v_y,v_z)(L_x/2-2,-y,L_z/2+z)~,
%    \end{equation}
  $z \rightarrow-z$ or shift-and-reflect (SR) 
  $(x,z)\rightarrow (x+L_x/2,-z)$ symmetry which both turn
  $(v_x,v_y,v_z)$ into $(v_x,v_y,-v_z)$ and $(b_x,b_y,b_z)$ into
  $(-b_x,-b_y,b_z)$. They also have either $z$-shift (S) 
  $z\rightarrow z+L_z/2$ or double-shift (DS) symmetry 
  $(x,z)\rightarrow(x+L_x/2,z+L_z/2)$ which both turn $(\vec{v},\vec{b})$
  into $(\vec{v},-\vec{b})$. Finally, owing to the invariance of PCF
  under $z$-rotations by $\pi$, all solutions have shift-and-rotate 
  $(x,y,z)\rightarrow (L_x/2-x,-y,z+L_z/2)$ symmetry, which turns 
  fields $(h_x,h_y,h_z)$ into $(-h_x,-h_y, h_z)$ (all transformations
  require appropriate $x$ and $z$ phase choices).
% %  If $\vec{v}$ has reflect symmetry,
% % $\vec{b}$ has shift-and-reflect symmetry and vice versa.
% % Both $\vec{v}$ and $\vec{b}$ have shift-and-rotate symmetry.
% % , one of the 
% % following symmetries is also present
% % : $\vec{b}$ can be represented by odd
% % Fourier components in $z$ and 
   The $\alpha$-dependence of the growth rates of several modes  
   for the 2D-3C configuration of Fig.~\ref{fig1} is plotted in Fig.~\ref{fig2}.
 A $(y,z)$ cut through the marginal (R,S) mode is shown in
 Fig.~\ref{fig3}. 

\begin{figure}[h]
\resizebox{\hsize}{!}{
\includegraphics[height=6.cm]{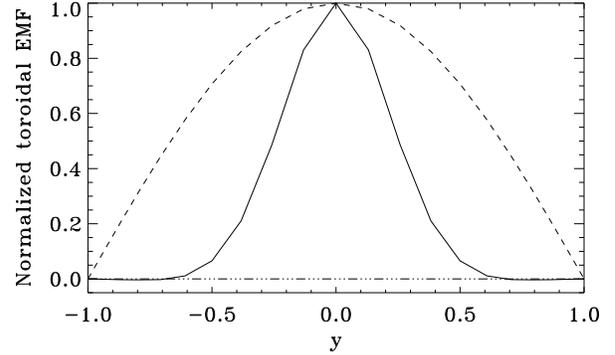}}
  \caption{Full line: normalized shearwise profile of the $k_z=\beta$
    component of the toroidal EMF created by the self-interactions of the
    three-dimensional $\alpha=0.375$ MRI mode of Fig.~\ref{fig3}.
    Dashed line: normalized shearwise profile of the artificial
    EMF~(\ref{eq:emfx}).}
  \label{fig4}
\end{figure}

\begin{figure}[h]
\resizebox{\hsize}{!}{
\includegraphics[height=6.cm]{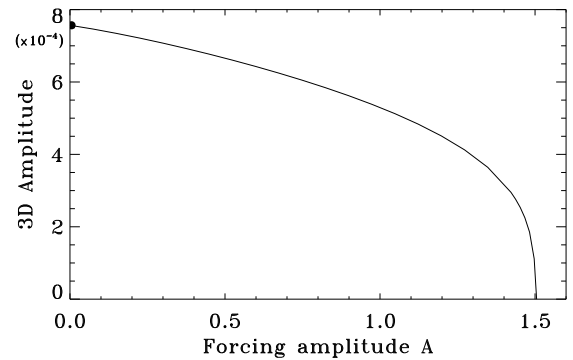}}
  \caption{$y$-integrated amplitude of the $k_x=0.375$, $k_z=0$ component of
    $v_y$ for  three-dimensional steady forced solutions, as a function of
    the forcing amplitude $A$. $(N_x,N_y,N_z)=(12,32,32)$.}% The  positive feedback of the MRI mode
%    (Fig.~\ref{fig4}) is responsible for the subcritical nature of the
%    bifurcation.}
  \label{fig5}
\end{figure}

\noindent In order to find a good initial guess to discover steady
three-dimensional solutions with the Newton solver, one 
    must select a marginally stable mode and check whether or not the
    $x$-averaged EMF due to its nonlinear interactions can
    take on the role of the artificial
EMF~(\ref{eq:emfx}) to sustain the poloidal field. This 
    depends strongly on the selected MRI mode and on its streamwise
 wavenumber $\alpha$, much like in the hydrodynamic problem \citep{wedin04}. If
the feedback is bad, the amplitude $A$ of the artificial EMF
    (and therefore the toroidal component of the
    2D-3C field) is adjusted and a 2D-3C flow marginally
    unstable to a MRI mode with slightly different
    $\alpha$ is recalculated. This way, we identify a
  (R,S) MRI eigenmode (Fig.~\ref{fig3}) whose nonlinear interactions
    create a toroidal EMF that has an interesting positive correlation
    with the artificial EMF (Fig.~\ref{fig4}). Note that
    considering the feedback in the 2D-3C momentum equation is not
    important  to find an interesting MRI mode since the goal
    is to replace an artificial forcing term imposed in
    the 2D-3C induction equation only.
%  Also,  the 2D-3C velocity field is weak and it
%     barely affects the computation  of the MRI modes.
\noindent We final attempt a continuation 
with respect to the forcing amplitude to find unforced ($A=0$)
solutions. The code is initialized with the 2D-3C
base flow of Fig.~\ref{fig1} plus a small amount of the three-dimensional
(R,S) MRI mode of Fig.~\ref{fig3}, and $A$ is set to a slightly smaller
value than that used to force the 2D-3C flow. 
The solver converges to a fully three-dimensional forced
solution, demonstrating that the bifurcation to three-dimensional
solutions is subcritical with respect to $A$.  Further confirmation of
the crucial role of the MRI mode feedback is obtained by performing
similar calculations initialized with marginal MRI modes with negative
feedback (such as the (SR,DS) mode of Fig.~\ref{fig2} at
$\alpha=0.355$), which reveal supercritical behaviour. 
The subcritical (R,S) branch can be followed down to $A=0$  (Fig.~\ref{fig5}),
at which point (black dot) nonlinear interactions due to the
$x$-dependent part of the solution fully take over the forcing term. 
The $A=0$ point is therefore a fully three-dimensional subcritical
nonlinear steady solution of the original MHD equations with no forcing
(Fig.~\ref{fig6}). The $x$-dependence of the solution, unlike for
$A=1.5$, can no longer be described by a single Fourier mode. 
  
\begin{figure}[h]
\resizebox{\hsize}{!}{
\includegraphics[height=6.cm]{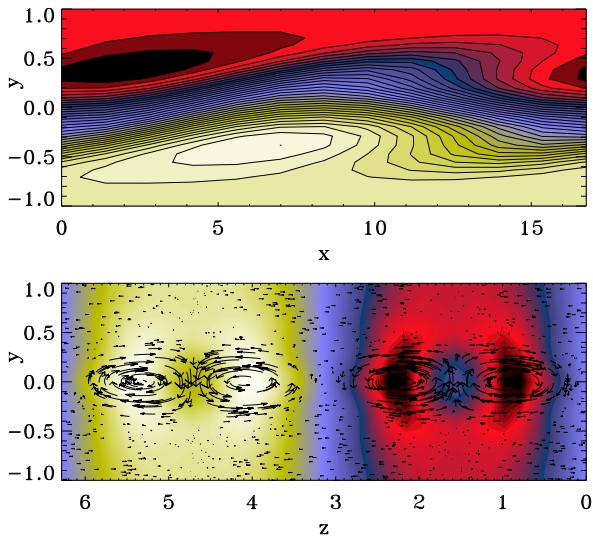}}
  \caption{Cuts through the unforced ($A=0$) nonlinear steady solution
    of Fig.~\ref{fig5} ($\alpha=0.375$, black dot). Top: 
    $b_z$ at $z=L_z/2$. Bottom: $b_x$ (color scale) and $(b_y,b_z)$ 
    (arrows) at $x=L_x/4$.}
  \label{fig6}
\end{figure}

% \begin{figure}[h]
% \resizebox{\hsize}{!}{
% \includegraphics[height=6.cm]{paramspace2.eps}}
%   \caption{Continuation of nonlinear dynamo solutions with respect to
%     $\rey$ for $\reym=1500$ (full line), $\reym=3000$ (dashed line)
%     and $(\alpha,\beta)=(0.375,1)$.}
%   \label{fig8}
% \end{figure}
As shown in Fig.~\ref{fig7}, this nonlinear steady solution
can be continued to large $\reym$ but exists only in
a narrow range of $\rey$. The reason for this could be that only at
low $\rey$ do marginally unstable MRI modes of the toroidal magnetic
field originate from a steady bifurcation. At larger $\rey$,  marginal
MRI modes systematically arise from a Hopf bifurcation instead:
pairs of steady eigenvalues corresponding to modes with the same
symmetry collide and turn into complex-conjugate pairs for
increasing $\rey$. 
%(see also Ref. \cite{matsumoto95} for similar observations in a
%slightly different problem).
To discover similar dynamo solutions at
larger $\rey$, it may therefore prove necessary to consider
time-dependent MRI modes instead of steady ones. This is
unfortunately numerically far more challenging, since the
symmetries used to reach decent three-dimensional
resolutions are broken when the MRI modes turn into
complex-conjugate pairs.

\begin{figure}[t]
\resizebox{\hsize}{!}{
\includegraphics[height=6.cm]{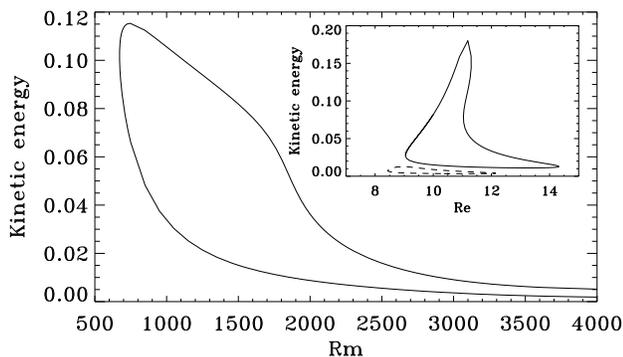}}
  \caption{Continuation with respect
    to $\reym$ for $\rey=10$, $(\alpha,\beta)=(0.375,1)$. Inset:
    continuation with respect to $\rey$ for $\reym=1500$
    (full line), $\reym=3000$ (dashed line). $(N_x,N_y,N_z)=(8,24,32)$.}
  \label{fig7}
\end{figure}

We have presented an instance of  self-sustaining, nonlinear dynamo solution in
Keplerian PCF, whose critical $\reym\simeq 670$ is comparable to that found 
in zero-net-flux simulations (another
$\reym$ definition is used in \cite{fleming00}). 
Preliminary direct spectral numerical simulations 
seem to confirm independently the existence of this solution, whose
structure is dominated by a coherent zero-net-flux toroidal magnetic field. 
% , but
% travelling motions should be considered to find solutions at large
% $\rey$ because marginally unstable MRI modes originate from a steady
% bifurcation only at low $\rey$.
We also discovered a (R,DS) branch, which makes it
probable that many solutions with different symmetries
exist.  Such coherent structures
%illustrate a very generic 
%mechanism for the sustenance of large-scale, zero net flux magnetic fields 
%in Keplerian shear flows.
are strictly speaking not turbulence but, like the hydrodynamics SSP, 
they probably act as organizing centers of the dynamics in
phase space \citep{waleffe98} and could play an important role in triggering and
sustaining MHD turbulence in magnetized Keplerian disks, where all the
basic ingredients of the dynamo are present.
% , the mechanism that we have been describing in this
% study represents a good candidate to explain the generation and sustenance
% of magnetic fields and MRI turbulence in these objects. 
Hopefully, this idealized study will be helpful to uncover the
details of the dynamo in more realistic set-ups.

% If this process can be extended to larger values of $\rey$ by in
% less constrained configurations and time-dependence
% permitted,  it is likely to trigger transition to MHD turbulence 
% in zero mean flux configurations. these results will certainly shed some
% new light on magnetic field generation and angular momentum transport in
% Keplerian accretion disks.

We thank C. Cossu for his contribution to the code
and A. Schekochihin and S.~Fromang for fruitful discussions.
We acknowledge support from the Leverhulme Trust and the Isaac Newton Trust.

\bibliographystyle{apsrev}
\bibliography{rincon}

\end{document}